\documentclass[twocolumn,showpacs,preprintnumbers,superscriptaddress,amsmath,amssymb,prl]{revtex4}

\usepackage{graphicx}
\usepackage{dcolumn}
\usepackage{bm}
\usepackage{color}

\begin{document}

\preprint{RCE Working Paper No. 2008-04}

\title{Multifractal detrended cross-correlation analysis for two nonstationary signals}

\author{Wei-Xing Zhou}
 \email{wxzhou@ecust.edu.cn}
 \affiliation{School of Business, East China University of Science and Technology, Shanghai 200237, China} %
 \affiliation{School of Science, East China University of Science and Technology, Shanghai 200237, China} %
 \affiliation{Research Center for Econophysics, East China University of Science and Technology, Shanghai 200237, China} %
 \affiliation{Research Center of Systems Engineering, East China University of Science and Technology, Shanghai 200237, China} %

\date{\today}

\begin{abstract}
It is ubiquitous in natural and social sciences that two variables,
recorded temporally or spatially in a complex system, are
cross-correlated and possess multifractal features. We propose a new
method called multifractal detrended cross-correlation analysis
(MF-DXA) to investigate the multifractal behaviors in the power-law
cross-correlations between two records in one or higher dimensions.
The method is validated with cross-correlated 1D and 2D binomial
measures and multifractal random walks. Application to two financial
time series is also illustrated.
\end{abstract}


\maketitle


Fractals and multifractals are ubiquitous in natural and social
sciences \cite{Mandelbrot-1983}.  The most usual records of
observable quantities in real world are in the form of time series
and their fractal and multifractal properties have been extensively
investigated. There are many methods proposed for this purpose
\cite{Taqqu-Teverovsky-Willinger-1995-Fractals,Montanari-Taqqu-Teverovsky-1999-MCM}.
For a single nonstationary time series, the detrended fluctuation
analysis (DFA) can be adopted to explore its long-range
autocorrelations
\cite{Peng-Buldyrev-Havlin-Simons-Stanley-Goldberger-1994-PRE,Hu-Ivanov-Chen-Carpena-Stanley-2001-PRE}
and multifractal features
\cite{Kantelhardt-Zschiegner-Bunde-Havlin-Bunde-Stanley-2002-PA}.
The DFA method can also be extended to investigate
higher-dimensional fractal and multifractal measures
\cite{Gu-Zhou-2006-PRE}.

There are many situations that several variables are simultaneously
recorded that exhibit long-range dependence or multifractal nature,
such as the velocity, temperature and concentration fields in
turbulent flows
\cite{Meneveau-Sreenivasan-Kailasnath-Fan-1990-PRA,Schmitt-Schertzer-Lovejoy-Brunet-1996-EPL,Beaulac-Mydlarski-2004-PF},
topographic indices and crop yield in agronomy
\cite{Kravchenko-Bullock-Boast-2000-AJ,Zeleke-Si-2004-AJ}, asset
prices, indexes, and trading volumes in financial markets
\cite{Ivanova-Ausloos-1999-EPJB,Matia-Ashkenazy-Stanley-2003-EPL}.
Recently, a first method named detrended cross-correlation analysis
(DXA) is proposed to investigate the long-range cross-correlations
between two nonstationary time series, which is a generalization of
the DFA method \cite{Podobnik-Stanley-2008-PRL}. Here we show that
the DXA method can be generalized to unveil the multifractal
features of two cross-correlated signals and higher-dimensional
multifractal measures. The validity and potential utility of the
multifractal detrended cross-correlation analysis (MF-DXA) method is
illustrated using one- and two-dimensional binomial measures,
multifractal random walks (MRWs), and financial prices.

Consider two time series $\{x_i\}$ and $\{y_i\}$, where $i=1, 2,
\cdots, M$. Without loss of generality, we can assume that these two
time series have zero means. Each time series is covered with
$M_s=[M/s]$ non-overlapping boxes of size $s$. The profile within
the $v$th box $[l_v+1,l_v+s]$, where $l_v=(v-1)s$, are determined to
be $X_v(k) = \sum_{j=1}^{k} x(l_v+j)$ and $Y_v(k) = \sum_{j=1}^{k}
y(l_v+j)$, $k=1,\cdots,s$. Assume that the local trends of
$\{X_v(k)\}$ and $\{Y_v(k)\}$ are $\{\widetilde{X}_v(k)\}$ and
$\{\widetilde{Y}_v(k)\}$, respectively. There are many different
methods for the determination of $\widetilde{X}_v$ and
$\widetilde{Y}_v$. The trend functions could be polynomials
\cite{Hu-Ivanov-Chen-Carpena-Stanley-2001-PRE}. The detrending
procedure can also be carried out nonparametrically based on the
empirical mode decomposition method
\cite{Wu-Huang-Long-Peng-2007-PNAS}. The detrended covariance of
each box is calculated as follows
\begin{equation}
 F_{v}(s) = \frac{1}{s}\sum_{k=1}^s \left[X_v(k)-\widetilde{X}_v(k)\right]\left[Y_v(k)-\widetilde{Y}_v(k)\right]
\end{equation}
The $q$th order detrended covariance is calculated as follows
\begin{equation}
 F_{xy}(q,s) = \left[\frac{1}{m}\sum_{v=1}^m F_{v}(s)^{q/2}\right]^{1/q}
\end{equation}
when $q\neq0$ and
\begin{equation}
 F_{xy}(0,s) = \exp\left[\frac{1}{2m}\sum_{v=1}^m \ln F_{v}(s)\right]~.
\end{equation}
We then expect the following scaling relation
\begin{equation}
 F_{xy}(q,s) \sim s^{h_{xy}(q)}~.
 \label{Eq:Fxy:s}
\end{equation}
When $X=Y$, the above method reduces to the classic multifractal
DFA.

\begin{figure}[htb]
\centering
\includegraphics[width=6.5cm]{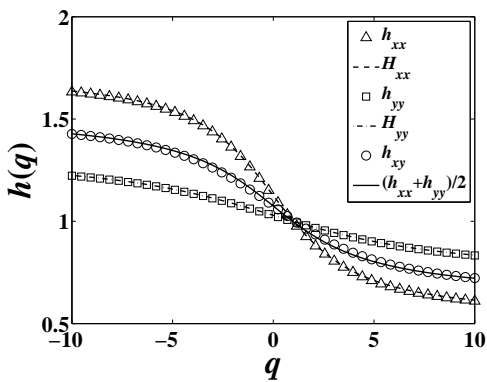}
\caption{\label{Fig:MFDXA:1D:pModel} Scaling exponents $h_{xy}(q)$
estimated using the multifractal detrended cross-correlation
analysis of two cross-correlated binomial measures generated from
the $p$-model. The numerically estimated exponents $h_{xx}(q)$ and
$h_{yy}(q)$ obtained from the multifractal detrended fluctuation
analysis of $x(i)$ and $y(i)$ are also presented, which well match
the analytical curves $H_{xx}(q)$ and $H_{yy}(q)$. This example
illustrates the relation
$h_{xy}(q)=\left[h_{xx}(q)+h_{yy}(q)\right]/2$.}
\end{figure}

In order to test the validity of the MF-DXA method, we construct two
binomial measures from the $p$-model with known analytic
multifractal properties as a first example
\cite{Meneveau-Sreenivasan-1987-PRL}. Each multifractal signal is
obtained in an iterative way. We start with the zeroth iteration $g
= 0$, where the data set $z(i)$ consists of one value, $z^{(0)}(1)=
1 $. In the $g$th iteration, the data set $\{z^{(g)}(i): i = 1, 2,
\cdots, 2^g\}$ is obtained from $z^{(g)}(2k+1)= pz^{(g-1)}(2k+1)$
and $z^{(g)}(2k)= (1-p)z^{(g-1)}(2k)$ for $k = 1, 2, \cdots,
2^{g-1}$. When $g\to\infty$, $z^{(g)}(i)$ approaches to a binomial
measure, whose scaling exponent function $h_{zz}(q)$ has an analytic
form
\cite{Halsey-Jensen-Kadanoff-Procaccia-Shraiman-1986-PRA,Meneveau-Sreenivasan-1987-PRL}
\begin{equation}
 H_{zz}(q) = 1/q-\log_2[p^q+(1-p)^q]/q~.
\end{equation}
In our simulation, we have performed $g=17$ iterations with $p_x =
0.3$ for $x(i)$ and $p_y = 0.4$ for $y(i)$. The cross-correlation
coefficient is 0.82. We find that $F_{xy}$, $F_{xx}$ and $F_{yy}$
all scale with respect to $s$ as power laws. Note that there are
evident log-periodic oscillations decorating power laws, which is an
inherent feature of the constructed binomial measures
\cite{Sornette-1998-PR}. The best estimates of the power-law
exponents are obtained when $s$ is sampled log-periodically
\cite{Zhou-Jiang-Sornette-2007-PA}. The resultant power-law
exponents $h_{xy}$, $h_{xx}$ and $h_{yy}$ are illustrated in
Fig.~\ref{Fig:MFDXA:1D:pModel}. The MF-DFA analysis gives
$h_{xx}(q)=H_{xx}(q)$ and $h_{yy}(q)=H_{yy}(q)$. We also find that
\begin{equation}
 h_{xy}(q) = [h_{xx}(q)+h_{yy}(q)]/2~.
 \label{Eq:MFDXA:hhh}
\end{equation}
For monofractal ARFIMA signals, this relation with $q=2$ is also
observed \cite{Podobnik-Stanley-2008-PRL}.

\begin{figure}[h!]
\centering
\includegraphics[width=6.5cm]{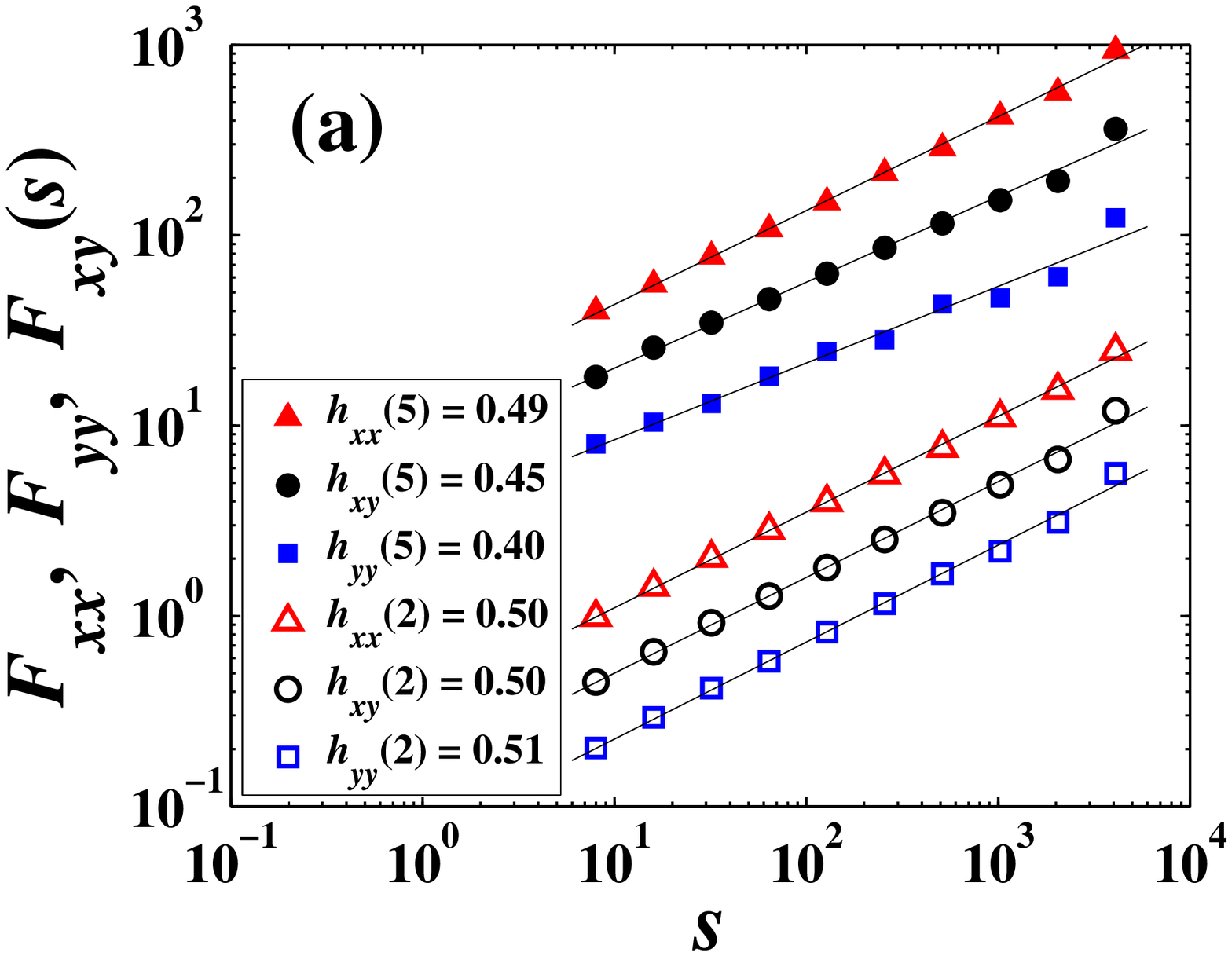}
\includegraphics[width=6.5cm]{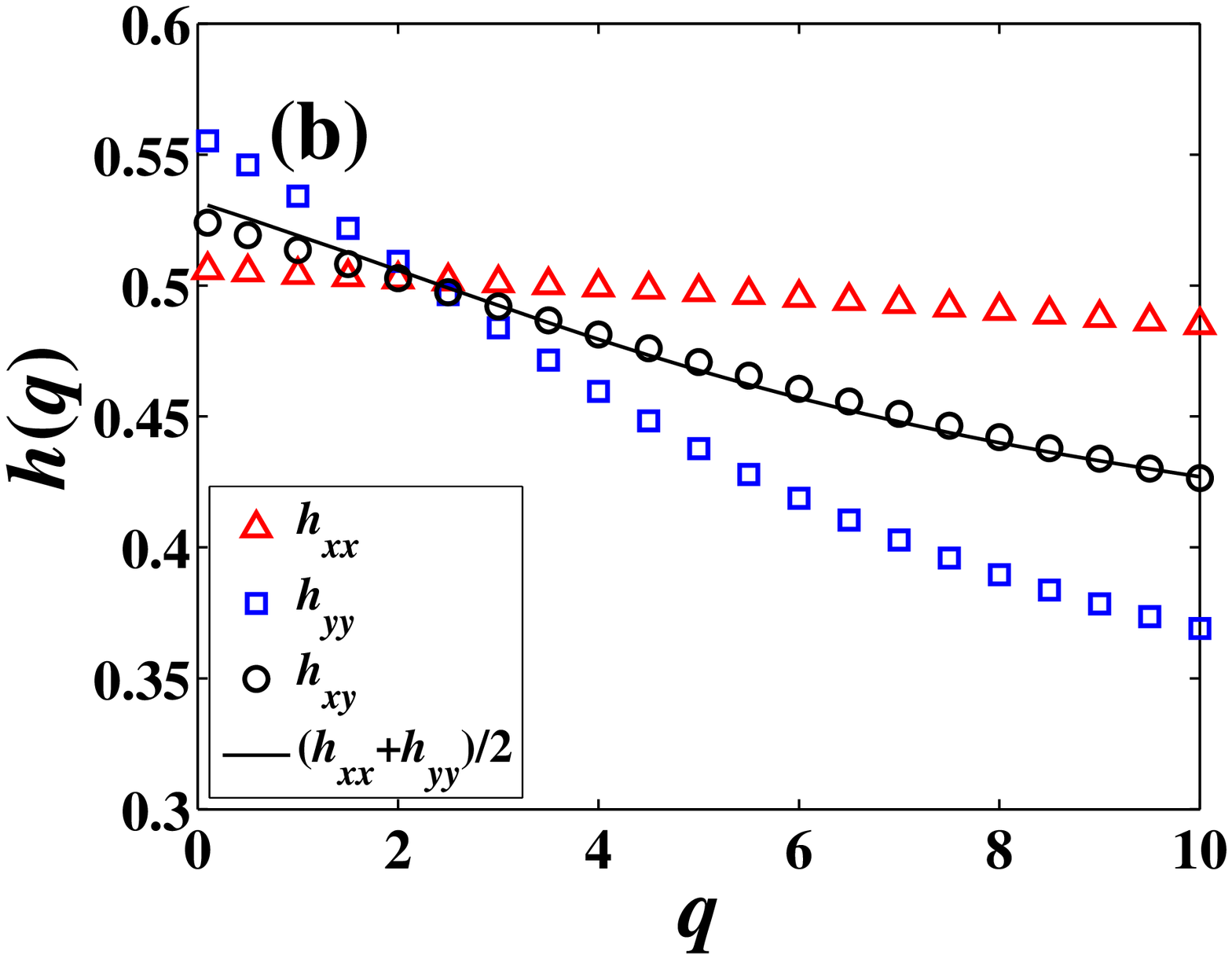}
\caption{\label{Fig:MFDXA:1D:MRW} (color online). Multifractal
nature in the power-law cross-correlations of two MRWs. (a)
Power-law scaling in $F_{xy}$, $F_{xx}$ and $F_{yy}$ with respect to
$s$ for $q=2$ and 5; (b) Power-law exponents $h_{xy}$, $h_{xx}$ and
$h_{yy}$.}
\end{figure}

As a second example, we consider the multifractal random walks
(MRWs) \cite{Bacry-Delour-Muzy-2001-PRE}. The increments of an MRW
are $\epsilon[k]e^{\omega[k]}$, where $\epsilon[k]$ and $\omega[k]$
are uncorrelated and $\omega[k]$ is a white noise. In order to
generated two cross-correlated MRWs, we can generate two time series
$\epsilon_x$ and $\epsilon_y$ possessing the properties in the MRW
formwork and rearrange $\epsilon_y$ such that the rearranged series
$\epsilon_y$ has the same rank ordering as $\epsilon_x$
\cite{Bogachev-Eichner-Bunde-2007-PRL}. We generate two MRW signals
of size $2^{16}$ with $\lambda^2=0.02$ for $x(i)$ and
$\lambda^2=0.04$ for $y(i)$, whose cross-correlation coefficient is
0.70. When $q$ is negative, no evident power-law scaling is observed
for $F_{xy}(s)$, which has great fluctuations. When $q$ is positive,
nice power-law scaling is observed for $F_{xy}$, $F_{xx}$ and
$F_{yy}$, as illustrated in Fig.~\ref{Fig:MFDXA:1D:MRW}(a) for $q=2$
and 5. The power-law exponents $h_{xy}$, $h_{xx}$ and $h_{yy}$ are
illustrated in Fig.~\ref{Fig:MFDXA:1D:MRW}(b). We see that
Eq.~(\ref{Eq:MFDXA:hhh}) holds in repeated numerical experiments.
However, this relation does not hold for some other realizations of
MRWs.

\begin{figure}[htb]
\centering
\includegraphics[width=6.5cm]{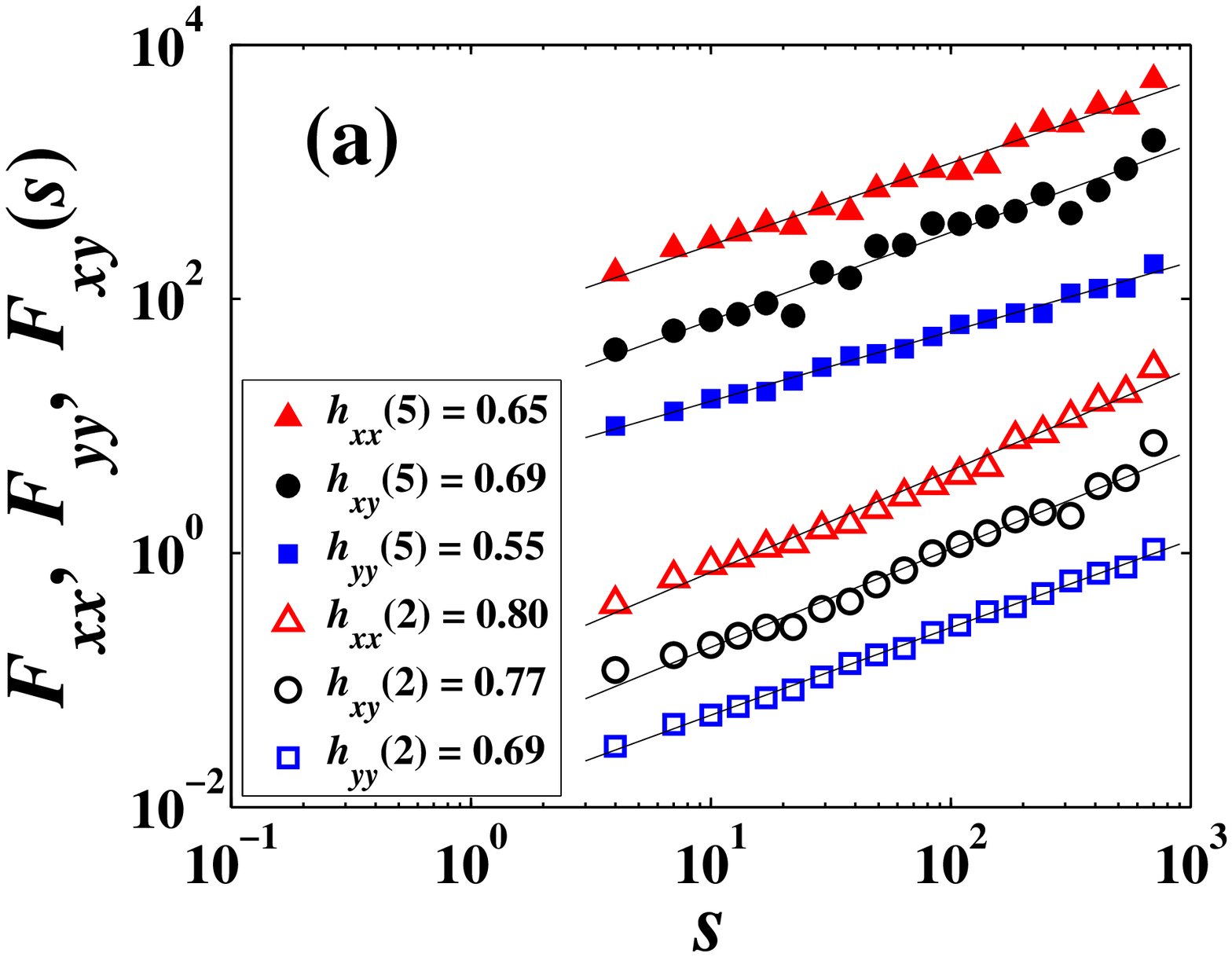}
\includegraphics[width=6.5cm]{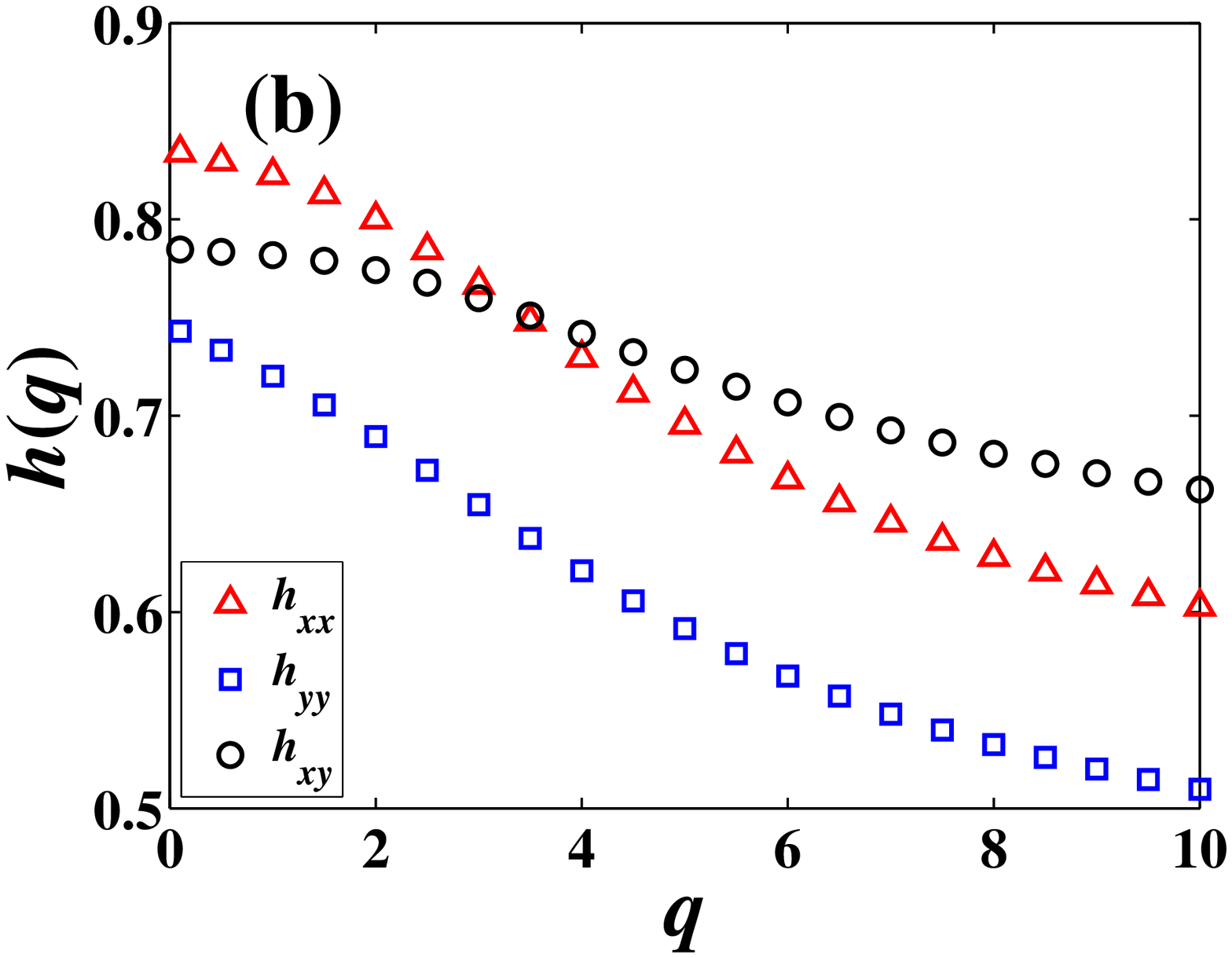}
\caption{\label{Fig:MFDXA:1D:DJIA:NASDAQ} (color online).
Multifractal nature in the power-law cross-correlations of the
absolute values of daily price changes for DJJIA and NASDAQ indexes
in the period from July 1993 to November 2003. (a) Power-law scaling
in $F_{xy}$, $F_{xx}$ and $F_{yy}$ with respect to $s$ for $q=2$ and
5. The scaling range is the same as in
Ref.~\cite{Podobnik-Stanley-2008-PRL}. (b) Dependence of the
power-law exponents $h_{xy}$, $h_{xx}$ and $h_{yy}$ as nonlinear
functions of $q$, indicating the presence of multifractality. There
is no clear relation between these exponents.}
\end{figure}

We now apply the MF-DXA method to the daily closing prices of DJIA
and NASDAQ indexes. For comparison, we have used the same data sets
and same scaling range as in Ref.~\cite{Podobnik-Stanley-2008-PRL}.
No evident power-law scaling is observed for negative $q$ values.
For positive $q$, we see power-law dependence of $F_{xy}$, $F_{xx}$
and $F_{yy}$ against time lag $s$. Two examples are illustrated in
Fig.~\ref{Fig:MFDXA:1D:DJIA:NASDAQ}(a) for $q=2$ and 5, where the
case of $q=2$ reproduces the results in
Ref.~\cite{Podobnik-Stanley-2008-PRL}. The power-law exponents
$h_{xy}$, $h_{xx}$ and $h_{yy}$ are depicted in
Fig.~\ref{Fig:MFDXA:1D:MRW}(b), which are nonlinear functions with
respect to $q$. We see that each time series of the absolute returns
possesses multifractal nature and their power-law cross-correlations
also exhibit multifractal nature.

We can generalize the 1D MF-DFA to the 2D version and its extension
to higher dimensions is straightforward. Consider two self-similar
(or self-affine) surfaces of identical sizes, which can be denoted
by two arrays $x(i,j)$ and $y(i,j)$, where $i=1,2,\cdots,M$, and $j
= 1, 2,\cdots, N$. The surfaces are partitioned into $M_s \times
N_s$ disjoint square segments of the same size $s\times s$, where
$M_s = [M/s]$ and $N_s = [N/s]$. Each segment can be denoted by
$x_{v,w}$ or $y_{v,w}$ such that $x_{v,w}(i,j)=x(l_v+i,l_w+j)$ and
$y_{v,w}(i,j)=y(l_v+i,l_w+j)$ for $1\leqslant{i,j}\leqslant{s}$,
where $l_v=(v-1)s$ and $l_w=(w-1)s$.

For each segment $x_{v,w}$ identified by $v$ and $w$, the cumulative
sum $X_{v,w}(i,j)$ is calculated as follows:
\begin{equation}
X_{v,w}(i,j) = \sum_{k_1=1}^{i}\sum_{k_2=1}^{j}{x_{v,w}(k_1,k_2)}~,
 \label{Eq:MFDXA:Xvw}
\end{equation}
where $1\leqslant{i,j}\leqslant{s}$. The cumulative sum
$Y_{v,w}(i,j)$ can be calculated similarly from $y_{v,w}$. The
detrended covariance of the two segments can be determined as
follows,
\begin{eqnarray}
F_{v,w}(s) = \frac{1}{s^2}\sum_{i = 1}^{s}\sum_{j =
1}^{s}\left[X_{v,w}(i,j)-\widetilde{X}_{v,w}(i,j)\right]\nonumber\\
\times\left[Y_{v,w}(i,j)-\widetilde{Y}_{v,w}(i,j)\right]~.
 \label{Eq:MFDXA:2D:Fvw}
\end{eqnarray}
where $\widetilde{X}_{v,w}$ and $\widetilde{Y}_{v,w}$ are the local
trends of $X_{v,w}$ and $Y_{v,w}$, respectively. The trend function
is pre-chosen in different function forms \cite{Gu-Zhou-2006-PRE}.
The simplest function could be a plane $\widetilde{u}(i,j) =
ai+bj+c$, which is adopted to test the validation of the method. The
overall detrended cross-correlation is calculated by averaging over
all the segments, that is,
\begin{equation}
F_{xy}(q,s) = \left\{\frac{1}{M_sN_s}\sum_{v = 1}^{M_s}\sum_{w =
1}^{N_s}{\left[F_{v,w}(s)\right]^{q/2}}\right\}^{1/q}~,
 \label{Eq:MFDXA:2D:Fq}
\end{equation}
where $q$ can take any real value except for $q = 0$. When $q = 0$,
we have
\begin{equation}
F_{xy}(q,s) = \exp\left\{\frac{1}{2M_sN_s}\sum_{v = 1}^{M_s}\sum_{w
= 1}^{N_s}{\ln[F_{v,w}(s)]}\right\}~,
 \label{Eq:MFDXA:2D:F0}
\end{equation}
according to L'H\^{o}spital's rule. The scaling relation between the
detrended fluctuation function $F_{xy}(q,s)$ and the size scale $s$
can be determined as follows
\begin{equation}
F_{xy}(q,s) \sim s^{h_{xy}(q)}~.
 \label{Eq:MFDXA:2D:Fqs}
\end{equation}
Since $N$ and $M$ need not be a multiple of the segment size $s$,
two orthogonal strips at the end of the profile may remain. Taking
these ending parts of the surface into consideration, the same
partitioning procedure can be repeated starting from the other three
corners \cite{Kantelhardt-Bunde-Rego-Havlin-Bunde-2001-PA}.

It is noteworthy to point out that the order of cumulative summation
and partitioning is crucial in the analysis of two- or
higher-dimensional multifractals. Consider the point located at
$(l_v+i,l_w+j)$ in the box identified by $v$ and $w$, where
$1\leqslant{i,j}\leqslant{s}$. The cumulative sum $X(l_v+i,l_w+j)$
can be expressed as follows
\begin{eqnarray}
 X(l_v+i,l_w+j)
 = X_{v,w}(i,j)
  +\sum_{k_1 = 1}^{l_v}    \sum_{k_2 = 1}^{l_w}    {x(k_1,k_2)}\nonumber\\
  +\sum_{k_1 = 1}^{l_v}    \sum_{k_2 = l_w}^{l_w+j}{x(k_1,k_2)}
  +\sum_{k_1 = l_v}^{l_v+i}\sum_{k_2 = 1}^{l_w}    {x(k_1,k_2)}~.
 \label{Eq:MFDXA:2D:Order}
\end{eqnarray}
For any pair of $(i,j)$, $X_{v,w}(i,j)$ is localized to the segment
$x_{v,w}$, while $X(l_v+i,l_w+j)$ contains extra information outside
the segment as shown above, which is not constant for different $i$
and $j$ and thus can not be removed by the detrending procedure. We
find that the power-law scaling is absent if $X(l_v+i,l_w+j)$ is
used in Eq.~(\ref{Eq:MFDXA:Xvw}). This observation is analogous to
the case of higher-dimensional detrended fluctuation analysis
\cite{Gu-Zhou-2006-PRE}.

We now present numerical experiments validating the two-dimensional
multifractal detrended cross-correlation analysis. There exist
several methods for the synthesis of two-dimensional multifractal
measures or multifractal rough surfaces
\cite{Decoster-Roux-Arneodo-2000-EPJB}. The most classic method
follows a multiplicative cascading process, which can be either
deterministic or stochastic
\cite{Mandelbrot-1974-JFM,Meneveau-Sreenivasan-1987-PRL,Novikov-1990-PFA,Meneveau-Sreenivasan-1991-JFM}.
The simplest one is the $p$ model proposed to mimic the kinetic
energy dissipation field in fully developed turbulence
\cite{Meneveau-Sreenivasan-1987-PRL}. Starting from a square, one
partitions it into four sub-squares of the same size and assigns
four given proportions of measure $p_{11}$, $p_{12}$, $p_{21}$, and
$p_{22}$ to them. Then each sub-square is divided into four smaller
squares and the measure is redistributed in the same way. This
procedure is repeated $g$ times and we generate multifractal
``surfaces'' of size $2^g \times 2^g$. The analytical expression of
$H_{xx}(q)$ or $H_{yy}(q)$ for individual multifractals is
\begin{equation}\label{Eq:MFDXA:2D:Hq}
    H(q) = \left[2-\log_2\left(p_{11}^q+p_{12}^q+p_{21}^q+p_{22}^q\right)\right]/q~.
\end{equation}

\begin{figure}[htb]
\centering
\includegraphics[width=6.5cm]{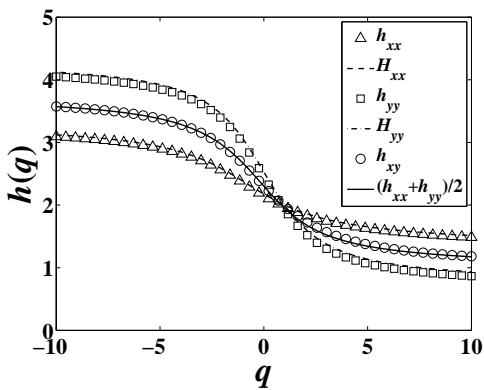}
\caption{\label{Fig:MFDXA:2D:pModel} Multifractal detrended
cross-correlation analysis of two cross-correlated synthetic
binomial measures from the $p$-model. The size of each multifractal
is $4096\times4096$ and the cross-correlation coefficient is 0.48.
The numerical exponents $h_{xx}(q)$ and $h_{yy}(q)$ obtained from
the multifractal detrended fluctuation analysis of $X$ and $Y$
locate approximately on the analytical curves $H_{xx}(q)$ and
$H_{yy}(q)$. This example illustrates the relation
$h_{xy}(q)=\left[h_{xx}(q)+h_{yy}(q)\right]/2$.}
\end{figure}

In our simulation, we have used $p_{11}=0.10$, $p_{12}=0.20$,
$p_{21}=0.30$, and $p_{22}=0.40$ for $X$ and $p_{11}=0.05$,
$p_{12}=0.15$, $p_{21}=0.20$, and $p_{22}=0.60$ for $Y$. We find
that the cross-correlation coefficient between the two multifractals
depends linearly on the generation number $g$: $c =
-0.0408g+0.9528$, where the value of R-squared is 0.9997. The 95\%
confidence intervals for the slope and intercept are  $[-0.0415,
-0.0402]$ and $[0.9489,0.9566]$, respectively. In our numerical
experiment, we have used $g=12$, which gives $c=0.48$. Very nice
power-law behaviors are confirmed in $F_{xy}(q,s)$, $F_{xx}(q,s)$,
and $F_{yy}(q,s)$ with respect to $s$ for different values of $q$.
The resultant power-law exponents $h_{xy}(q)$, $h_{xx}(q)$ and
$h_{yy}(q)$ are illustrated in Fig.~\ref{Fig:MFDXA:2D:pModel} marked
with open circles, squares and triangles, respectively. We find that
the relation $h_{xy}(q)=\left[h_{xx}(q)+h_{yy}(q)\right]/2$ holds.

In summary, we have proposed a multifractal detrended
cross-correlation analysis to explore the multifractal behaviors in
power-law cross-correlations between two simultaneously recorded
time series or higher-dimensional signals. The MF-DXA method is a
combination of multifractal analysis and detrended cross-correlation
analysis. Potential fields of application include turbulence,
financial markets, ecology, physiology, geophysics, and so on.

\begin{acknowledgments}
We thank J.-F. Muzy for help in generating two cross-correlated MRWs
and G.-F. Gu for discussions. This work was partly supported by NSFC
(70501011), Fok Ying Tong Education Foundation (101086), Shanghai
Rising-Star Program (06QA14015), and Program for New Century
Excellent Talents in University (NCET-07-0288).
\end{acknowledgments}

\bibliography{E:/papers/Auxiliary/Bibliography}

\end{document}